\documentclass{rmf-d}
\usepackage{nopageno,rmfbib,multicol,times,epsf,amsmath,amssymb,cite}
\usepackage[latin1]{inputenc}
\usepackage[]{caption2}
\usepackage{graphicx}
\usepackage{hyperref}
\usepackage{lineno}
\usepackage{color}
\def\rmfcornisa{Research OR Education of Physics \hfill\rmf\ {\bf ??} (*?*) ???--???
\hfill MES? A\~NO?}

\def\rmfcintilla{{\it Rev.\ Mex.\ Fis.\/} {\bf ??} (*?*) (????) ???--???}
\clearpage \rmfcaptionstyle 
\begin{document}
\title{   The $Z_{cs}(3985)^-$ structure in a coupled-channels model
\vspace{-6pt}}
\author{ Pablo G. Ortega     }
\address{ Departamento de F\'isica Fundamental and Instituto Universitario de F\'isica
Fundamental y Matem\'aticas (IUFFyM), Universidad de Salamanca, E-37008 Salamanca, Spain  }
\author{ David R. Entem and Francisco Fern\'andez }
\address{ Grupo de F\'isica Nuclear and Instituto Universitario de F\'isica
Fundamental y Matem\'aticas (IUFFyM), Universidad de Salamanca, E-37008
Salamanca, Spain}
\maketitle
\recibido{day month year}{day month year
\vspace{-12pt}}
\begin{abstract}
\vspace{1em}
The discovery of the $Z_c(3900)^\pm$ and $Z_b(10610)^\pm$ structures in the heavy quarkonium spectrum showed the need to incorporate  hadron structures beyond the naive $qqq$ and $q\bar q$ systems in quark models.
The new charged structure called $Z_{cs}(3985)^-$, spotted in the $K^+$ recoil-mass spectrum close to the  $D^-_s D^{*0}/D^{*-}_sD^0$ threshold, is a new
evidence in this line.

In this work, we analyze the $Z_{cs}(3985)^-$ state, following the calculation of the $Z_c$ and $Z_b$ states using a chiral constituent quark model in a coupled-channels calculation, with all the parameters constrained from previous calculations.
The pole structure of the S-matrix shows two virtual poles below the $D_s^-D^{*\,0}$ and $D_s^{*\,-}D^{*\,0}$ thresholds compatible with the $Z_{cs}(3985)^\pm$ and a new predicted $Z_{cs}(4110)^\pm$ structure, the SU(3) flavor partner of the $Z_c(4020)^\pm$.
The $K^+$ recoil-mass spectrum is calculated in good agreement with LHCb and BESIII experimental data, with no fine tuning of the model parameters.
Our results indicate that the $Z_{cs}(3985)^\pm$ and $Z_{cs}(4000)^\pm$ signals originate from the same virtual state.
 \vspace{1em}
\end{abstract}
\keys{  Potential models, Charmed strange mesons, Exotic mesons  \vspace{-4pt}}
\pacs{   \bf{\textit{12.39.Pn, 14.40.Lb, 14.40.Rt}}    \vspace{-4pt}}
\begin{multicols}{2}

\section{Introduction} \label{sec:introduction}

Recently, the BESIII Collaboration has announced the discovery of a new structure, dubbed $Z_{cs}(3985)^-$, in the $K^+$ recoil-mass spectrum in the process $e^+e^-\rightarrow K^+(D_s^-D^{*0}+D_s^{*-}D^0)$ at $\sqrt{s}=4.681$ GeV~\cite{BESIII:2020qkh}. Its pole mass and width, setted with a mass-dependent width Breit-Wigner line shape, are

\begin{align}
M^{\rm pole}_{\rm Z_{cs}}&=(3982.5^{+1.8}_{-2.6}\pm 2.1) {\rm MeV}/c^2, \\
\Gamma^{\rm pole}_{\rm Z_{cs}}&= (12.8^{+5.3}_{-4.4}\pm 3.0) {\rm MeV},
\end{align}
being the first uncertainties statistical and the second ones systematic.

The minimum quark content of the $Z_{cs}(3985)^-$ is most likely $c\bar c s\bar u$, which is explicitly exotic and resembles other charge meson-like structures such as the $Z_c(3900)^\pm$/$Z_c(4020)^\pm$ in the charmonium~\cite{BESIII:2013ris} or the $Z_b(10610)^\pm$/$Z_b(10650)^\pm$ in the bottomonium spectrum~\cite{Belle:2011aa}.

As those analogous structures, the new $Z_{cs}(3985)^-$ is close to a meson-meson threshold, the $D_s^-D^{*0}$ and $D_s^{*-}D^0$ thresholds. Its mass is about $100$ MeV/$c^2$ larger than that of the $Z_c(3900)^\pm$, which is the typical mass difference between $D^{(*)}_s$ and $D^{(*)}$ mesons. These two characteristics suggest, on the one hand, that the $Z_{cs}(3985)^-$ may have a significant molecular $D_s^-D^{*0}+D_s^{*-}D^0$ component and, on the other hand, that it could be the $SU(3)$ flavor partner of the $Z_c(3900)^\pm$, where a $u$ or $d$ quark is replaced by a $s$ quark in its quark content.

Few months later, the LHCb Collaboration also announced the observation of two $J^P=1^+$ $c\bar c u\bar s$ exotic states decaying to $J/\psi K^+$ final channels, in the reaction $B^+\to J/\psi\phi K^+$~\cite{LHCb:2021uow}: The so-called $Z_{cs}(4000)^+$ resonance, with a mass of $4003\pm 6^{+4}_{-14}$ MeV/$c^2$ and a width of $131\pm15\pm26$ MeV, and the $Z_{cs}(4220)^+$, with a mass of $4216\pm 24^{+43}_{-30}$ MeV/$c^2$ and a width of $233\pm52^{+97}_{-73}$ MeV. Based on their experimental analysis, no evidence was found that supported that the $Z_{cs}(4000)^+$ is the same as the $Z_{cs}(3985)^-$ observed by BESIII, though their energy proximity suggest a relation between both structures which is worth exploring.

The theoretical interpretations of such states include QCD sum rules~\cite{Wan:2020oxt,Wang:2020rcx,Wang:2020iqt,Azizi:2020zyq,Xu:2020evn,Albuquerque:2021tqd,Ozdem:2021yvo}, one boson exchange models~\cite{Chen:2020yvq}, quarks models~\cite{Jin:2020yjn} and effective field theories~\cite{Yang:2020nrt,Meng:2020ihj}. Most of the calculations reproduced the mass and, in some cases, the width of the $Z_{cs}(3985)^-$ as a $D_s^-D^{*0}+D_s^{*-}D^0$ $I(J^{P})=\frac{1}{2}(1^{+})$ molecular state.

In this work we investigate the structure of the $Z_{cs}(3985)^-$ structure in the framework of a widely-tested constituent quark model, to analyze if it can be understood as the $SU(3)_F$ partner of the $Z_c(3900)^\pm$~\cite{Ortega:2018cnm}.
For that purpose, a coupled-channels calculation of the $J^P=1^+$ charm-strange sector is performed in order to describe the $Z_{cs}(3985)^-$, using the same chiral constituent quark model (CQM) and including the following channels: $D^{*0}D_s^-$, $D^0 D_s^{*-}$, $D^{*0} D_s^{*-}$, $J/\psi K^{*\,-}(892)$, $\eta_c K^{*\,-}(892)$ and $J/\psi K^-$. For more details on the present calculation, the reader is referred to Ref.~\cite{Ortega:2021enc}.

\section{Theory}
\label{sec:theory}

For the description of the $Z_{cs}(3985)^-$ we employ the same chiral constituent quark model (CQM) used to describe the $Z_c(3900)^\pm/Z_c(4020)^\pm$ and $Z_b(10610)^\pm/Z_b(10650)^\pm$ structures~\cite{Ortega:2018cnm,Ortega:2021xst}. Details of the model, explicit expressions of the potentials and model parameters can be found in, e.g., Ref.~\cite{Vijande:2004he,Segovia:2008zz}.

The CQM assumes that the QCD phenomenology can be described by the following effective Lagrangian at low-energy~\cite{Diakonov:2002fq}
\begin{equation}
{\mathcal L} = \bar{\psi}(i\, {\slash\!\!\! \partial} -M(q^{2})U^{\gamma_{5}})\,\psi  \,,
\end{equation}
which proposes that quarks acquire a dynamical momentum dependent mass $M=M(q^2)$, with $M(q^2\to\infty)=m_q$, as a consequence of the spontaneous breaking of the chiral symmetry at some momentum scale, and the subsequent emergence of Goldstone-boson exchange interactions between light quarks.

This is modeled in $U^{\gamma_5} = e^{i\lambda _{a}\phi ^{a}\gamma _{5}/f_{\pi}}$, the matrix of Goldstone-boson fields, where $f_\pi$ is the pion decay constant, $\lambda^a$ are the $SU(3)$ colour matrices and $\phi^a$ denotes the pseudoscalar fields ($\vec \pi$,$K_i$,$\eta_8$), with $i=1,\ldots,4$.
This matrix of Goldstone-boson fields can be expanded as
\begin{equation}
U^{\gamma _{5}} = 1 + \frac{i}{f_{\pi}} \gamma^{5} \lambda_{a} \phi^{a} - \frac{1}{2f_{\pi}^{2}} \phi_{a} \phi^{a} + \ldots
\end{equation}
The constituent quark mass is obtained from the first term, the second one describes the pseudoscalar meson-exchange interaction among quarks and the main contribution of the third term comes from the two-pion exchange, which is modeled by means of a scalar-meson exchange potential. Goldstone-boson exchanges are considered when the two quarks are light, but they do not appear in the other two configurations: light-heavy and heavy-heavy.

QCD perturbative effects appear beyond the chiral-symmetry breaking scale, which are included in the CQM
by means of the one-gluon exchange potential, derived from the following vertex Lagrangian term
\begin{equation}
{\mathcal L}_{qqg} = i\sqrt{4\pi\alpha_{s}} \, \bar{\psi} \gamma_{\mu}
G^{\mu}_a \lambda^a \psi,\label{Lqqg}
\end{equation}
where $\alpha_{s}$ is the strong coupling constant and $G^{\mu}_a$ is the gluon field. The strong coupling constant, $\alpha_{s}$, has a scale dependence which allows a consistent description of light, strange and heavy mesons.

Quark confinement is modeled through a linear potential screened at large inter-quark distances, due to
sea quarks effect~\cite{Bali:2005fu}:
\begin{equation}
V_{\rm CON}(\vec{r}\,)=\left[-a_{c}(1-e^{-\mu_{c}r})+\Delta \right]  (\vec{\lambda}_{q}^{c}\cdot\vec{\lambda}_{\bar{q}}^{c}) \,.
\label{eq:conf}
\end{equation}
Here, $a_{c}$ and $\mu_{c}$ are model parameters. One can see that the potential is linear at short inter-quark distances with an effective confinement strength $\sigma = -a_{c} \, \mu_{c} \, (\vec{\lambda}^{c}_{i}\cdot \vec{\lambda}^{c}_{j})$, while it becomes constant at large distances.
The one-gluon exchange and confining potentials are flavor-blind in our CQM.

Besides the direct t-channel interaction mediated by Goldstone bosons, quarks and antiquarks can interact through s-channel processes, dubbed 'annihilation' processes for now on. In the case of the $c\bar c s\bar u$ quark content, the annihilation proceeds through the one-gluon and one-kaon exchanges for the $c\bar c$ and $s\bar u$ pairs, respectively.

With the aim of describing the interaction among $D$ and $D_s$ mesons, we employ the Resonating Group Method~\cite{Wheeler:1937zza}.
That way, we are able to connect the interaction at meson level, considered as quark-antiquark clusters, with the microscopic interaction among constituent quarks from the CQM.

We assume that the wave function of a system composed of two mesons $A$ and $B$ with distinguishable quarks can be written as~\footnote{Note that, for simplicity of the discussion presented herein, we have dropped off the spin-isospin wave function, the product of the two color singlets and the wave function that describes the center-of-mass
motion.}
\begin{equation}
\langle \vec{p}_{A} \vec{p}_{B} \vec{P} \vec{P}_{\rm c.m.} | \psi
\rangle = \phi_{A}(\vec{p}_{A}) \phi_{B}(\vec{p}_{B})
\chi_{\alpha}(\vec{P}) \,,
\label{eq:wf}
\end{equation}
where  $\phi_{C}(\vec{p}_{C})$ is the wave function of a general meson $C$ with $\vec{p}_{C}$ the relative momentum between the quark and antiquark of the meson $C$, calculated by means of the two-body
Schr\"odinger equation using the Gaussian Expansion Method~\cite{Hiyama:2003cu}. The wave function which takes into account the relative motion of the two mesons is $\chi_\alpha(\vec{P})$.

The projected Schr\"odinger equation for the relative wave function can be written as follows:
\begin{align}
&
\left(\frac{\vec{P}^{\prime 2}}{2\mu}-E \right) \chi_\alpha(\vec{P}') + \sum_{\alpha'}\int \Bigg[ {}^{\rm RGM}V_{D}^{\alpha\alpha'}(\vec{P}',\vec{P}_{i}) + \nonumber \\
&
+ {}^{\rm RGM}V_{R}^{\alpha\alpha'}(\vec{P}',\vec{P}_{i}) \Bigg] \chi_{\alpha'}(\vec{P}_{i})\, d\vec{P}_{i} = 0 \,,
\label{eq:Schrodinger}
\end{align}
where $E$ is the energy of the system. The direct potential ${}^{\rm RGM}V_{D}^{\alpha\alpha '}(\vec{P}',\vec{P}_{i})$ can be written as
\begin{align}
&
{}^{\rm RGM}V_{D}^{\alpha\alpha '}(\vec{P}',\vec{P}_{i}) = \sum_{i\in A, j\in B} \int d\vec{p}_{A'} d\vec{p}_{B'} d\vec{p}_{A} d\vec{p}_{B} \times \nonumber \\
&
\times \phi_{A}^{\ast}(\vec{p}_{A'}) \phi_{B}^{\ast}(\vec{p}_{B'})
V_{ij}^{\alpha\alpha '}(\vec{P}',\vec{P}_{i}) \phi_{A'}(\vec{p}_{A}) \phi_{B'}(\vec{p}_{B})  \,.
\end{align}
The quark rearrangement potential ${}^{\rm RGM}V_{R}^{\alpha\alpha'}(\vec{P}',\vec{P}_{i})$ represents a natural way to connect meson-meson channels with different quark content, such as $ J/\psi K^-$ and $D^{*\,0}D_s^-$, and it is given by
\begin{align}
&
{}^{\rm RGM}V_{R}^{\alpha\alpha'}(\vec{P}',\vec{P}_{i}) = \sum_{i\in A, j\in B} \int d\vec{p}_{A'}
d\vec{p}_{B'} d\vec{p}_{A} d\vec{p}_{B} d\vec{P} \phi_{A}^{\ast}(\vec{p}_{A'}) \times \nonumber \\
&
\times  \phi_{B}^{\ast}(\vec{p}_{B'})
V_{ij}^{\alpha\alpha '}(\vec{P}',\vec{P}) P_{mn} \left[\phi_{A'}(\vec{p}_{A}) \phi_{B'}(\vec{p}_{B}) \delta^{(3)}(\vec{P}-\vec{P}_{i}) \right] \,,
\label{eq:Kernel}
\end{align}
where $P_{mn}$ is the operator that exchanges quarks between clusters.

The solution of the coupled-channels RGM equations is performed deriving from Eq.~\eqref{eq:Schrodinger} a set of coupled Lippmann-Schwinger equations of the form
\begin{align}
T_{\alpha}^{\alpha'}(E;p',p) &= V_{\alpha}^{\alpha'}(p',p) + \sum_{\alpha''} \int
dp''\, p^{\prime\prime2}\, V_{\alpha''}^{\alpha'}(p',p'') \nonumber \\
&
\times \frac{1}{E-{\cal E}_{\alpha''}(p^{''})}\, T_{\alpha}^{\alpha''}(E;p'',p) \,,
\end{align}
where $\alpha$ labels the set of quantum numbers needed to uniquely define a certain partial wave, $V_{\alpha}^{\alpha'}(p',p)$ is the projected potential that contains the direct and rearrangement potentials, and ${\cal E}_{\alpha''}(p'')$ is the energy corresponding to a momentum $p''$, written in the nonrelativistic case as:
\begin{equation}
{\cal E}_{\alpha}(p) = \frac{p^2}{2\mu_{\alpha}} + \Delta M_{\alpha} \,.
\end{equation}

Here, $\mu_{\alpha}$ is the reduced mass of the $AB$ system corresponding to the channel $\alpha$, and $\Delta M_{\alpha}$ is the difference between the threshold of the $AB$ system and the one we take as a reference.

Once the $T$-matrix is calculated, we determine the on-shell part which is directly related to the scattering matrix (in the case of nonrelativistic kinematics):
\begin{equation}
S_{\alpha}^{\alpha'} = 1 - 2\pi i
\sqrt{\mu_{\alpha}\mu_{\alpha'}k_{\alpha}k_{\alpha'}} \,
T_{\alpha}^{\alpha'}(E+i0^{+};k_{\alpha'},k_{\alpha}) \,,
\end{equation}
with $k_{\alpha}$ the on-shell momentum for channel $\alpha$.

Our aim is to explore the existence of states above and below thresholds within the same formalism. Thus, we have to analytically continue all the potentials and kernels for complex momenta in order to find the poles of the $S$-matrix in any possible Riemann sheet.

The $Z_{cs}(3985)^-$ has been discovered at BESIII in the $K^+$ recoil-mass spectrum of the $e^+e^-\to K^+(D_s^-D^{*\,0}+D_s^{*\,-}D^0)$ process at $\sqrt{s}$ ranging from $4.628$ to $4.698$ GeV~\cite{BESIII:2020qkh}~\footnote{For simplicity we only discuss the $e^+e^-$ production. However, similar expressions can be employed for the production in $pp$ collisions as measured by LHCb, with a different parametrization of the vertex.}. The line shapes of the $Z_{cs}(3985)^-$ structure will be calculated through the Lorentz-invariant production amplitude, ${\cal M}$, which describes the $Z_{cs}\to AB$ reaction and can be written as
\begin{align}\label{eq:amplitude}
\mathcal{M^\beta}&=\left({\cal A}^\beta e^{i\,\theta_\beta}-\sum_{\beta'}{\cal A}^{\beta'}e^{i\,\theta_{\beta'}} \int d^3p\frac{T^{\beta'\beta}(p,k^\beta,E)}{p^2/2\mu-E-i\,0}\right).
\end{align}
where $A^\beta$ and $\theta_\beta$ are parameters that describe the production amplitude and phase of the $(AB)_\beta$ channel from the $e^+e^-$ vertex.

From this amplitude, the line shape of a $e^+e^-\to K^+Z_{cs}$ reaction from a point-like vertex at a given $\sqrt{s}$ and the subsequent $Z_{cs}\to AB$ is expressed as

\begin{equation}
\frac{d\Gamma_{Z_{cs}\to AB}}{dm_{AB}} = \frac{1}{(2\pi)^3}\frac{k_{AB} k_{K Z_{cs}}}{4\,s}|{\cal M}^\beta(m_{AB})|^2  \,,
\end{equation}
with $\beta$ the quantum numbers of the channel $AB$, $m_{AB}$ is the invariant mass of the $AB$ meson pair and where $k_{KZ_{cs}}$ and $k_{AB}$ are the on-shell momentum of the $KZ_{cs}$ and $AB$ pairs, respectively.

A global normalization factor will be added to describe the experimental events measured in the $\sigma(e^+e^-\to K^+ Z_{cs}^-)\times \mathcal{B}(Z_{cs}^- \to D_s^-D^{*0}+D_s^{*-}D^0)$ process:
\begin{equation}
N(m_{AB}) = \mathcal{N}_{AB}\times \frac{d\Gamma_{Z_c\to AB}}{dm_{AB}} \,,
\end{equation}
which encodes other relevant process details such as the value of $\sigma(e^+e^-\to  K^+ Z_{cs}^-)$.

The set of  $\{{\cal A}_{AB},\theta_{AB},\mathcal{N}_{AB}\}$ parameters (for $AB$ the channels involved in the calculation) and their uncertainties are, then, obtained by means of a global $\chi^2$ function minimization procedure using the available experimental data on $D_s^-D^{*0}+D_s^{*-}D^0$.

\section{Results}
\label{sec:results}

In order to describe the $Z_{cs}(3985)^-$ structure, we perform a coupled-channels calculation of the $I(J^{P})=\frac{1}{2}(1^{+})$ four quark sector. The included channels are those close to the experimental mass of the $Z_{cs}(3985)^-$ in a relative $S$-wave, that is: $J/\psi K^-$ (3592 MeV/$c^2$), $\eta_c K^{*\,-}$ (3877 MeV/$c^2$),   $D^-_s D^{*0}$ (3976 MeV/$c^2$), $D^0 D_s^{*\,-}$ (3979 MeV/$c^2$), $J/\psi K^{*\,-}$ (3990 MeV/$c^2$) and $D^{*0} D_s^{*-}$ (4120 MeV/$c^2$) channels, where the threshold energies are shown in parenthesis.

In the $D^{(*)}D_s^{(*)}$, the pion interaction is not allowed. Thus, in order to explore the effect of non-diagonal coupling among the $D^{(*)}D_s^{(*)}$ channels, we explored two different calculations, one without annihilation diagrams (model $a$) and another including annihilation diagrams (model $b$). Whereas in model $a$ the $D^{(*)}D_s^{(*)}$ are decoupled, in model $b$ annihilation diagrams provide a way to connect them.

From the analysis of the $S$-matrix, two poles are found below the $D^-_s D^{*0}$ and $D^{*0} D_s^{*-}$ thresholds in the second Riemann sheet at $3970$ MeV/$c^2$ and $4110$ MeV/$c^2$ for the model $a$ and  $(3961-3\it i)$ MeV/$c^2$ and $(4106 - 5\it i)$ MeV/$c^2$ for the model $b$. The first one would create the $Z_{cs}(3985)^-$ peak, seen as an enhancement above the $D^-_s D^{*0}+D^0 D_s^{*\,-}$ thresholds, whereas the second one would be an unseen $Z_{cs}^-$ state, denoted as $Z_{cs}(4110)^-$, analog of the $Z_c(4020)^\pm$ in the charm-strange sector, which could give rise to the recent $Z_{cs}(4220)^+$ structure discovered by LHCb~\cite{LHCb:2021uow}.

In our model,  both the $Z_{cs}(3985)^-$ and the predicted $Z_{cs}(4110)^-$ are not resonances but virtual state and, therefore, a direct comparison with the complex energy of the pole of a Breit-Wigner parametrization of a resonance would be misleading. That is the reason why the best way to confront our results with the experimental data is through the description of the line shapes.

\begin{center}
	\includegraphics[width=\columnwidth]{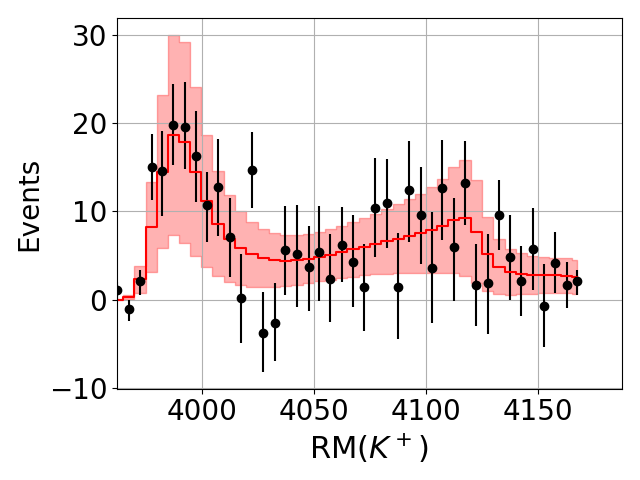} \\
	\includegraphics[width=\columnwidth]{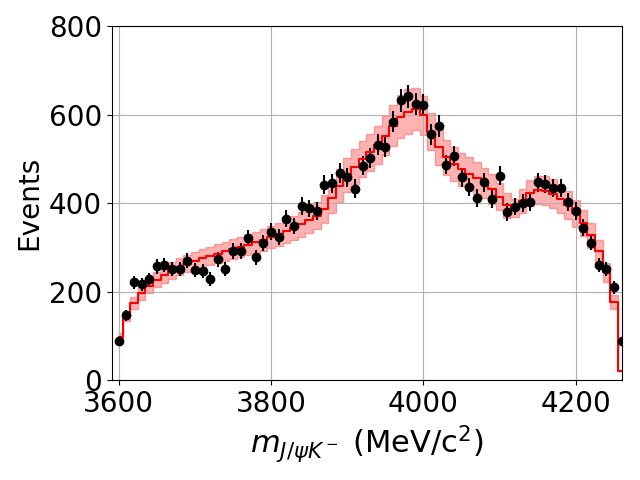} \\

	\refstepcounter{figure}
	\rmfcaptionstyle{Figure \ref{fig1}: Theoretical description (solid lines) of the experimental $K^+$ recoil-mass spectra (black dots) measured by BESIII~\cite{BESIII:2020qkh} (upper panel) and of the $J/\psi K^-$ invariant mass spectrum measured by LHCb~\cite{LHCb:2021uow} (lower panel) for {\it model b}. The red shadowed-area around the line represents the 68\% CL of the fit. We remark here that the fit only affects the production vertex, with no fine-tuning of the CQM parameters in the description of the coupled-channels S-matrix. }\label{fig1} \\
\end{center}

Both models, $a$ and $b$, provide a good agreement with the experimental line shapes, with similar accuracy, so the coupling among $D_s^{(*)}D^{(*)}$ does not appear to be relevant to reproduce the experimental data. Hence, we will only show results for model $b$ (see Ref.~\cite{Ortega:2021enc} for more details and further discussion). Our results for the $K^+$ recoil-mass spectra are shown in the upper panel of Fig.~\ref{fig1}. The shaded area around the theoretical curve shows the statistical 68\%-confident level (CL) of the fit, obtained by propagating the errors of the fitted parameters by means of the covariance matrix. The values for the normalization factors and amplitudes are shown in Table~\ref{tab:norm}, the result on the $\chi^2/{\rm d.o.f.}$ is also collected therein.

Our model reproduces the experimental data without any fine-tuning of its parameters, the same ones used to describe the $Z_c(3900)^\pm$ and the $Z_c(4020)^\pm$ states, besides the unavoidable normalization and amplitude factors that describe the inner details of the production vertex, which involves further dynamics not relevant here.

\tabletopline\vspace{2pt}\lilahf{\sc Table I.\ {\rm \label{tab:norm} Vertex parameters for the $D^-_s D^{*0}+D_s^{*-}D^0$ (left) and $J/\psi K^-$ (right) line shapes for model $b$. The minimum value of the $\chi^2/{\rm d.o.f.}$, calculated in the $[3.9,4.2]$ GeV energy range, is also given. The 68\% uncertainty in the parameters, in parenthesis, is obtained from the fit.}}
\begin{center}
\small{\renewcommand{\arraystretch}{1.3}
\renewcommand{\tabcolsep}{1.35pc}
 \begin{tabular}{ccc}
 \hline
  Parameters & BESIII data & LHCb data \\
  \hline
$\chi^2/{\rm d.o.f.}$  		&	1.02		&	2.04	\\
$\ln\left({\cal N}_{D_sD^*+DD_s^*}\right)$	&	24.6(6)	&	-	\\
$\ln\left({\cal N}_{J/\psi K}\right)$ 		&	-	&	25.43(8)\\
${\cal A}_{J/\psi K}$ 	&	1.0(9)	&	0.028(2)\\
${\cal A}_{\eta_c K^*}$ &	0.33(2)	&	0.35(3)	\\
${\cal A}_{D_sD^*}$ 	&	0.052(5)&	0.02(1)	\\
${\cal A}_{DD_s^*}$ 	&	0.04(1)	&	0.10(1)	\\
${\cal A}_{J/\psi K^*}$ &	0.01(2)	&	0.52(7)	\\
${\cal A}_{D^*D_s^*}$ 	&	0.29(2)	&	0.15(1)	\\
$\theta_{J/\psi K}$ 	&	-2.65(8)&	-0.39(13)\\
$\theta_{\eta_c K^*}$ 	&	-1.8(2)	&	-3.19(6)\\
$\theta_{D_sD^*}$ 	&  	3.2(4)	&		-1.25(3)\\
$\theta_{DD_s^*}$ 	&  	2.4(2)	&		-0.96(11)\\
$\theta_{J/\psi K^*}$ 	&	-3(5)	&	1.15(13)\\
$\theta_{D^*D_s^*}$		&	-2.7(2)	&	2.80(11)\\
  \hline
\end{tabular}}
\end{center}

The two virtual peaks showed by the theoretical calculation are compatible with the experimental states detected by BESIII and LHCb.
As the reaction involved in BESIII and LHCb are of different nature ($e^+e^-$ vs $pp$ collisions) and the center of mass energy is not the same, we do not expect that the same parameters that describe the vertex for BESIII hold for the LHCb data. Thus, a second fit on LHCb data was perform, whose results are shown in the lower panel of Fig.~\ref{fig1}. Our results support the hypothesis that the $Z_{cs}(3985)^-$ and the $Z_{cs}(4000)^+$ are the same state, and that the $Z_{cs}(4220)^+$ is an effect of an event dip around the $D^{*0} D_s^{*-}$ threshold, which emerges as a second peak in the $D^-_s D^{*0}+D_s^{*-}D^0$ line shape, around the same energy.

Finally, we can use the same framework to predict states in the hidden bottom strange sector. Doing an analogous coupled-channels calculation in this sector we obtain two virtual poles below the $B^{*\,-}B_s^0$ and $B^{*\,-}B_s^{*\,0}$ thresholds at $10691$ MeV/$c^2$ and $10739$ MeV/$c^2$, respectively. These states should be identified as the $SU(3)_F$ partners of the $Z_b(10610)^\pm$ and the $Z_b(10650)^\pm$, heavy partners of the $Z_{cs}(3985)^-$ and $Z_{cs}(4110)^-$, and could be detected in the $\Upsilon(1S)K^-$ and $B^{*-} B^0_s+B^{-} B^{*0}_s$ channels.


\section{Acknowledgments}
This work has been funded by
Ministerio de Ciencia e Innovaci\'on
under Contract No. PID2019-105439GB-C22/AEI/10.13039/501100011033
and by EU Horizon 2020 research and innovation program, STRONG-2020 project, under grant agreement No 824093.

\end{multicols}
\medline
\begin{multicols}{2}

\end{multicols}
\end{document}